\documentclass[pra,aps,twocolumn,superscriptaddress]{revtex4-2}

\usepackage{graphicx,amsmath,color,soul}

\usepackage[colorlinks,urlcolor=blue,citecolor=blue,linkcolor=blue]{hyperref}

\begin{document}

\title{Switching dynamics of femtosecond solitons in parity-time-symmetric coupled optical waveguides}

\author{Ambaresh Sahoo}
\email{sahoo.ambaresh@gmail.com}
\affiliation{Department of Physics, Indian Institute of Technology Guwahati, Assam 781039, India}

\author{Dipti Kanika Mahato}
\affiliation{Department of Physics, Indian Institute of Technology Guwahati, Assam 781039, India}

\author{A. Govindarajan}
\affiliation{Department of Nonlinear Dynamics, School of Physics,
Bharathidasan University, Tiruchirapalli 620024, India}

\author{Amarendra K. Sarma}
\email{aksarma@iitg.ac.in}
\affiliation{Department of Physics, Indian Institute of Technology Guwahati, Assam 781039, India}

\begin{abstract}

We report a detailed study on soliton steering dynamics in a parity-time-symmetric directional coupler in the femtosecond domain, which requires incorporation of higher-order perturbative effects such as  third-order and fourth-order dispersions, self-steepening, and intrapulse Raman scattering. With a high gain/loss, the combination of all these effects is found to stabilize the soliton pulse evolution in the coupler from the chaotic behavior of unperturbed evolution. This work demonstrates that efficient soliton steering can be achieved at very low critical power and a relatively higher gain/loss even in the femtosecond regime.

\end{abstract}

\maketitle

\section{Introduction}

In the domains of high-speed signal processing and ultrafast communication,  all-optical switching devices are considered to be the key elements that have drawn significant attention over the past few decades \cite{Sasikala18}. In this context, a dual-core nonlinear directional  coupler featuring power-dependent switching has been studied extensively \cite{Trillo88, friberg1987ultrafast, friberg1988femotosecond}. In recent times, the first experimental observation of the parity-time-symmetric ($\mathcal{PT}$-symmetric) effect \cite{Ruter10} in  a linear waveguide directional coupler with balanced gain/loss has provided an opportunity for further research \cite{El-Ganainy18,suchkov2016nonlinear,suneera2019switching}. In this connection, several studies have shown that such couplers are beneficial for all-optical switching when operating in the nonlinear regime \cite{sukhorukov2010nonlinear, ramezani2010unidirectional, dmitriev2011scattering}. Furthermore, the stability of  optical solitons and their dynamical control by periodic management in $\mathcal{PT}$-symmetric fiber couplers have been investigated semi-analytically and numerically \cite{alexeeva2012optical, Driben11,Fan19}.

Recently, it has been reported that the requirement of high critical power for optical solitons can be reduced sharply by introducing $\mathcal{PT}$-symmetry in a nonlinear directional coupler \cite{Govindarajan19}. However, the study deals with the system with unperturbed coupled nonlinear Schrodinger equation (NLSE),  whose domain of operation is practically limited to the picosecond (ps) timescale only with the moderate value of gain/loss. In the domain of ultra-short bright solitons ranging from few femtosecond (fs) to 1 ps, the practical application of any real fiber coupler requires incorporation of higher-order perturbative effects such as higher-order dispersions (HODs) including third-order dispersion (TOD) and fourth-order dispersion (FOD), self-steepening (SS), and intrapulse Raman scattering (IRS) \cite{Blow89,GPAbook} in the coupled NLSE. In another work, we have investigated all-optical switching of bistable solitons that emerge in non-Kerr type media with saturable nonlinearity \cite{Sahoo22}, where we characterize the switching dynamics in the presence of $\mathcal{PT}$-symmetric configurations in the ps timescale. However, for ultrafast switching in the fs time scale, one has to exploit Kerr solitons where various higher-order effects may appear depending on the nonlinear media.

In this work, it is of particular interest to explore the idea of $\mathcal{PT}$-symmetric fiber coupler for the fs pulse switching and observe the switching dynamics under the effect of balanced gain and loss. We first explore the effects of individual higher-order perturbations (IRS, SS, HODs) on switching dynamics. Following that,  we identify the suitable parameter region for the stable, efficient, and low power switching  assisted by the combined effects of all higher-order perturbations, which is helpful for the practical realization of the all-optical fs switching devices. It is to be mentioned that in the usual single NLSE case, the higher-order perturbations degrade the device performances. Also, the conventional coupler (i.e., without $\mathcal{PT}$ symmetry) could not be used in the context of fs switching, as it shows poor performance in terms of efficiency with multiple switching. In an earlier work \cite{Malomed97}, it has been reported that in a conventional coupler IRS restabilizes the symmetric solitons at sufficiently large energies. However, in the case of a $\mathcal{PT}$-symmetric coupler, we show that a partly radiating solitons caused by high gain/loss values are stabilized in the presence of higher-order perturbations. 
\vspace{-0.2cm}

\section{Model: $\mathcal{PT}$-symmetric fiber coupler}

\begin{figure}[t]
\centering
\begin{center}
\includegraphics[width=0.49\textwidth]{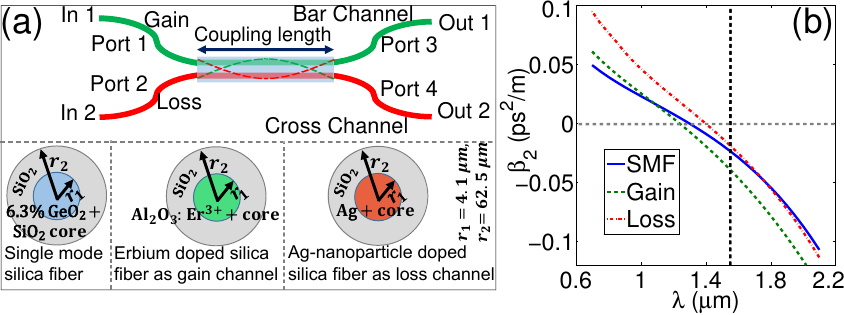}
\caption{(a) Schematic diagrams of a fiber coupler, with passive and active ports are depicted in the inset. (b) The GVD profiles of the three configurations are depicted with vertical dashed line representing the launching wavelength ($\lambda_0=1.55\,\mu$m).}\label{Schem}
\end{center}
\end{figure}
To discuss the fs pulse switching, we consider a realistic single-mode fiber coupler configuration [Fig.\,\ref{Schem}(a)] with an erbium-doped gain channel \cite{Myslinski97} and a silver-nanoparticle-doped loss channel \cite{Sahoo21}, the parameters of which can be adjusted by changing the doping concentrations to achieve a balanced gain/loss system ($\mathcal{PT}$ symmetry). Since the volume fraction of doping is small in both cases, the group-velocity dispersion (GVD) profiles show similar nature, as shown in Fig.\,\ref{Schem}(b). We also numerically confirm that the gain-dispersion does not have a significant role on the switching dynamics, as the switching power and efficiency are dependent on the total integral power. Based on these, we consider the generalized NLSE with an integral form of the nonlinearities \cite{GPAbook}. The generalized coupled-mode equations of the slowly-varying envelops $A_{1,2}(z,t)$ in the two channels of the fiber couplers made up of non-identical fibers (shown in Fig.\,\ref{Schem}) with respective normalized GVD and HOD parameters, $\delta_{n\,1,2}$, can be written in normalized units \cite{Govindarajan19}:
\begin{align} \label{eq:1}
&i { \partial_\xi u_{1,2} } +\sum _{ n=2 }^{ \infty  }{ { \delta  }_{ n \,1,2 }{\left(i { { \partial_\tau  }} \right)}^{ n }u_{1,2}  } \mp i\Gamma u_{1,2} +\kappa u_{2,1} \nonumber\\ &\hspace{1.9cm}  +\left(1+is { { \partial_\tau  }}\right)  R(\tau){\boldsymbol{\otimes}}|u_{1,2}|^{ 2 } u_{1,2} =0,
\end{align} 
where $u_{1,2}(\xi,\tau)=A_{1,2}/\sqrt{P_0}$, with $P_0$ being the peak power of the input pulse. The propagation distance ($z$) and time ($t$) variables are, respectively, normalized with new parameters as $\xi=z/L_D$ and $\tau=(t-zv_g^{-1})/t_0$, with $L_D=t_0^2/|\beta_2(\omega_0)|$, $t_0$ is the input pulse width, $v_g$ is the group-velocity of the pulse, and $\beta_2(\omega_0)$ being the GVD parameter at the carrier frequency $\omega_0$. The inter-core linear coupling ($K$) and balanced gain/loss ($G$) parameters are scaled as $\kappa=K L_D$ and $\Gamma=G L_D$.
Also, the terms $\delta_{n}\,[=\beta_{n}/(n!|\beta_2|t_0^{n-2})]$, $s\,[=1/(\omega_0t_0)]$, and $R(\tau)=(1-f_R)\delta(\tau)+f_R h_R(\tau)$ respectively denote the normalized parameters of HOD, SS or shock, and the nonlinear response function including Kerr nonlinearity via $\delta(\tau)$ and Raman response function $h_R(\tau)$ that connects through the convolution integration ${\boldsymbol{\otimes}}$ with the field envelops. Here, $h_R(\tau) = (1- f_b)(\tau_1^{-2} +\tau_2^{-2})\tau_1 \exp(-\tau/\tau_2) \sin(\tau/\tau_1)+f_b[(2\tau_b -\tau)/\tau_b^2]\exp(-\tau/\tau_b)$ \cite{Lin06}, with $f_R=0.245$ being the fractional contribution of the delayed Raman response to nonlinear polarization, $\tau_1 = 12.2\,{\rm fs}/t_0$, $\tau_2 = 32\,{\rm fs}/t_0$, $\tau_b\approx 96\,{\rm fs}/t_0$, and the relative contribution of the boson peak is included through $f_b = 0.21$. It is to be noted that, in principle, Eq.\,(\ref{eq:1}) should include different $v_g$ for non-identical fiber channels. However, for the fiber structures shown in Fig.\,\ref{Schem}(a), we calculate the group velocities at the carrier frequency $\omega_0(=2\pi c/\lambda_0)$: $v_g\approx1.99\times 10^8$\,m/s for the gain channel and $v_g\approx 2.01\times 10^8$\,m/s for the loss channel, which are almost identical. In order to keep our analysis simpler and to focus on the impact of higher-order perturbations on switching dynamics, we consider the same $v_g$ for two channels of the $\mathcal{PT}$-symmetric coupler and use the same co-moving pulse evolution equation [Eq.\,(\ref{eq:1})].

The system described by Eq.\,(\ref{eq:1}) is $\mathcal{PT}$-symmetric due to the presence of equal gain and loss ($\Gamma$) in the two waveguides. According to the non-Hermitian photonics, there exist three possible states for a $\mathcal{PT}$-symmetric system: unbroken ($\kappa>\Gamma$), broken ($\kappa < \Gamma$), and an exceptional point ($\kappa = \Gamma$) that indicates the phase-transition regime of $\mathcal{PT}$- symmetry \cite{El-Ganainy18}. Earlier study in the context of $\mathcal{PT}$- symmetric soliton switching has confirmed that a $2\pi$ coupler with the configuration shown in Fig.\,\ref{Schem}(a) exhibits richer steering dynamics, achieving low critical steering power ($P_{\rm cr}$) while maintaining excellent transmission efficiency \cite{Govindarajan19}. In view of this, we confine our analysis to the $2\pi$ $\mathcal{PT}$ coupler. Also, we restrict our system to work in the unbroken regime alone by fixing a condition of $\kappa>\Gamma$. This is due to the fact that power controlled steering is limited within the unbroken $\mathcal{PT}$- symmetric regime as the soliton pulse exhibits severe instability in the broken $\mathcal{PT}$-symmetric regime. Note that,  Eq.\,(\ref{eq:1}) contains all the higher-order terms that act as perturbations in the context of ultrashort (fs) pulse dynamics \cite{GPAbook}. When the pulse duration is large enough (ps or larger), IRS and SS can be ignored. Also, if the input pulse is launched far away from the zero-dispersion frequency, HODs can be neglected. In such a case,  Eq.\,(\ref{eq:1}) resembles the usual coupled-mode equation for $\mathcal{PT}$-symmetric systems, without higher-order perturbations \cite{Driben11, alexeeva2012optical, Govindarajan19}. To investigate the soliton switching dynamics, we solve the Eq.\,(\ref{eq:1}) taking into account all of the perturbations present in the system by launching the exact Kerr soliton solution of the unperturbed NLSE, $u_1(0,\tau) = \sqrt{\widetilde{P}}\, sech(\tau)$ and $u_2(0,\tau) = 0$, where $\widetilde{P}=t_0^2\gamma P_0/|\beta_2|$ is the dimensionless input peak power, with $\gamma\approx2.156\times10^{-3}$\,W$^{-1}$m$^{-1}$ being the nonlinear coefficient of the fiber at $\lambda_0=1.55$\,$\mu$m (which is approximately equal for both channels for low doping concentration of fibers). It is worth noting that the nonreciprocity behavior of $\mathcal{PT}$-symmetric systems is exploited here, which possess better transmission efficiency when a pulse is launched into the gain channel. The reverse (launching into the loss channel) is not efficient in this $\mathcal{PT}$ case, in contrast to the conventional coupler that follows reciprocity behavior. When both the channels are simultaneously excited, the pulse evolution becomes unstable and chaotic. This particular scheme can be utilized as a phase-sensitive switching, provided the power in the loss channel is very small compared to the gain channel \cite{Sahoo22}.
The numerical simulations are carried out using the split-step fast-Fourier transformation for coupled-mode equations \cite{GPAbook} incorporated with the fourth-order Runge-Kutta algorithm. The numerical investigations show that HODs and higher-order nonlinear phenomena have a significant impact on $\mathcal{PT}$- symmetric couplers. The study further reveals that for an ultrashort pulse propagation, the critical switching power $P_{cr}$ depends on $\delta_{n >2}$ (HOD parameters), IRS, and $s$ (SS or shock effect), i.e., $P_{\rm cr} \sim f(\delta_n,{\rm IRS}, s)$. The strength of these perturbations can be tuned by varying the input pulse duration $t_0$. We extensively study the pulse dynamics for a wide range of $t_0$ (from 5 fs to 1 ps). In order to get into the detail switching dynamics, we first systematically analyze the individual perturbation effects. Then, the combined effects of all the higher-order perturbations are discussed in some details.

\section{ IMPACT OF HIGHER-ORDER NONLINEAR EFFECTS}
\begin{figure}[t]
\centering
\begin{center}
\includegraphics[width=0.48\textwidth]{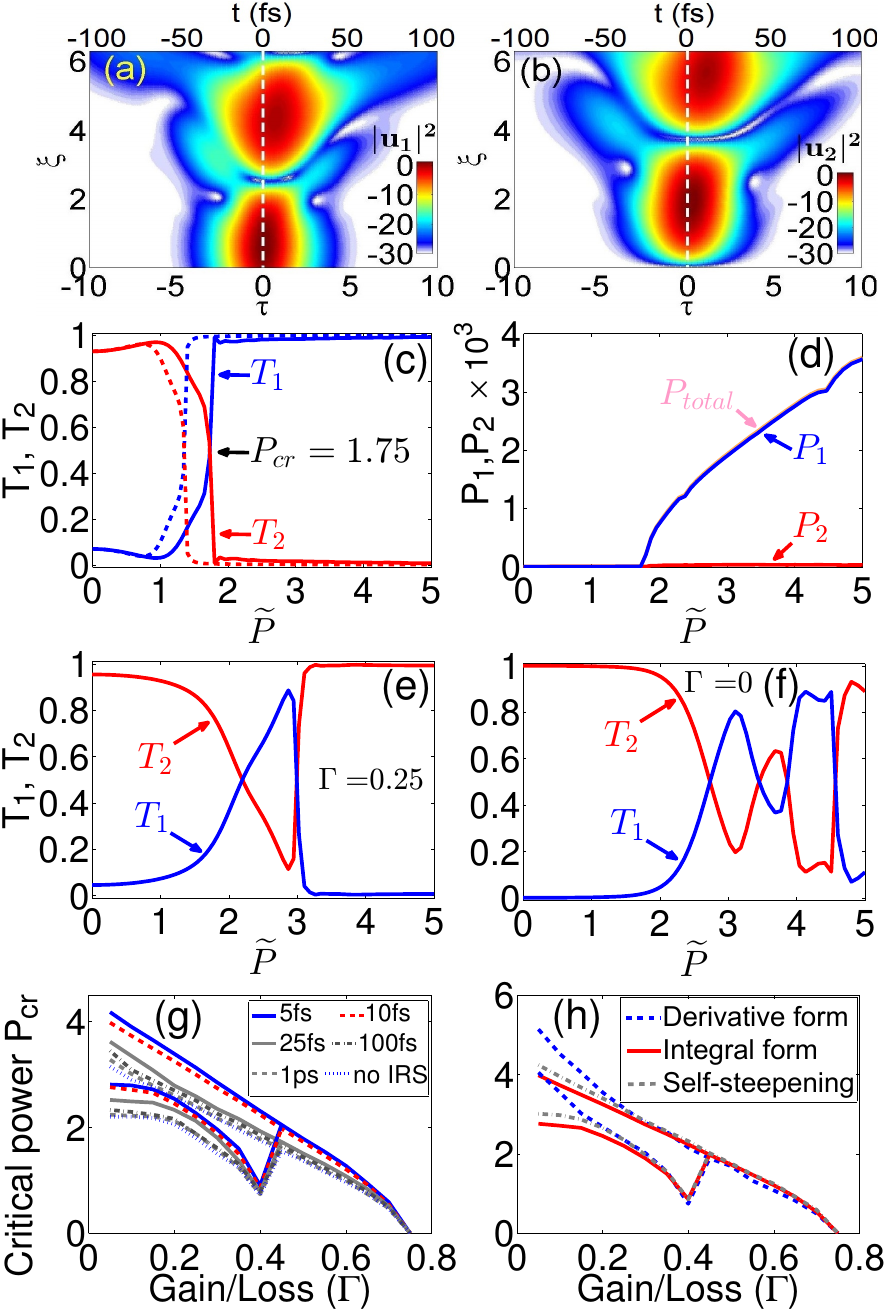}
\caption{(a), (b) The evolution of a 10\,fs $sech$ pulse in the two channels of a $2\pi$ $\mathcal{PT}$-symmetric fiber coupler with $\widetilde{P}=1$, $\kappa=1$, and $\Gamma=0.5$ under the effect of IRS only (in dB scale). The corresponding switching dynamics are depicted in (c) and (d) by solid curves. The dashed curves in (c) represent the unperturbed case [Eq.\,(\ref{eq:1}) without HOD, IRS, and SS]. (e) Bistable nature of switching dynamics for $\Gamma<0.45$ and (f) multiple switching of a conventional coupler ($\Gamma=0$) are shown with IRS perturbation. (g) Critical switching power $P_{\rm cr}$ vs $\Gamma$ as a function of $t_0$, and (h) relative comparison between the integral NLSE model and the derivative model for IRS perturbation with $t_0 = 10$ fs. The effect of SS on $P_{\rm cr}$ vs $\Gamma$ is also shown in (h).}\label{Raman}
\end{center}
\end{figure}
Higher-order perturbations on ultrashort pulses have been studied in various optical waveguides, and the effects on the pulse dynamics are significant. Ultrashort optical soliton in fibers with intense peak power can excite higher-order nonlinear effects like SS and IRS. Under the influence of IRS, the central frequency of the soliton experiences a redshift \cite{Gordon86}. IRS also influences the temporal dynamics of the soliton by imposing a temporal deceleration. In a single soliton case, the effect of IRS leads to Raman amplification and lasing. SS, on the other hand, creates an optical shock on the leading edge of the pulse, resulting in asymmetric spectral broadening \cite{GPAbook}. In order to investigate the effect of IRS and SS on the soliton steering dynamics in a $2\pi$ $\mathcal{PT}$-symmetric fiber coupler, we first numerically solve Eq.\,(\ref{eq:1}) for a $10\,$fs input \textit{sech} pulse in the presence of IRS only. The temporal evolutions are plotted in Figs.\,\ref{Raman}(a) and \ref{Raman}(b) for the two channels. Here, the power of a fs soliton launched into the first channel steers back and forth between the two channels and eventually exit from the second. The IRS-induced characteristic temporal decelerations are also evident, which differ from the unperturbed case represented by vertical dashed lines at the middle. From our study on the spatial evolution of powers $P_{1,2}=\int_{-\infty}^{\infty}|u_{1,2}(L_c,\tau)|^2d\tau$ with $L_c$ being the coupling length and transmission coefficients $T_{1,2}=P_{1,2}/(P_1 + P_2)$ of the two channels in the coupler (not shown here), we find that the IRS induces the soliton to acquire an additional relative phase difference than unperturbed case \cite{Govindarajan19}. For a short optical pulse, in the absence of any perturbation, inclusion of $\mathcal{PT}$-symmetry revealed a critical switching power to be $P_{\rm cr} = 1.34$ with a sharp transmission efficiency [dotted curve in Fig.\,\ref{Raman}(c)]. However, for a pulse with $t_0 = 10\,$fs, the IRS perturbation present in the system increases the critical switching power to $P_{\rm cr} = 1.75$ while maintaining the transmission efficiency at nearly 99$\%$ as acheived in the unperturbed case \cite{Govindarajan19}. As a result, above the critical switching power, almost all of the output energy will be captured by the first channel, with a negligible amount by the second channel, as shown in Fig.\,\ref{Raman}(d). The transmission characteristics for the $2\pi$ coupler with $\Gamma=0.25$ and $\Gamma=0$ (conventional coupler) with IRS perturbation are also depicted in Figs.\,\ref{Raman}(e) and \ref{Raman}(f), respectively. Here the $\mathcal{PT}$-coupler ($\Gamma=0.25$) strictly shows double switching dynamics leading to a two distinct threshold power bellow a critical $\Gamma$ ($<0.45$), a characteristic feature of $2\pi$ $\mathcal{PT}$-symmetric couplers after which the two switching cycles mingle with a  single threshold power and so the total power remains in the first channel [shown in Fig.\,\ref{Raman}(d)]. In contrast, the conventional coupler manifests in less efficient and unwanted multiple steering curves.  
Additionally, we explore the role of the input pulse duration $t_0$ in order to demonstrate the relationship between critical switching power and the gain/loss parameter in a $2\pi$ $\mathcal{PT}$-coupler.  Figure\,\ref{Raman}(g) demonstrates that with the decrease in the strength of IRS parameter (as $t_0$ increases), the value of the critical switching power ($P_{\rm cr}$) also reduces accordingly, and vice versa. This suggests that the loss/gain parameter $\Gamma$, beyond $0.6$ is more appropriate for effective fs switching as it gives rise to a very low critical power, which is not possible in the ps regime. Following that, for the completeness of the study, we perform a numerical analysis with the derivative form of the IRS perturbation $[f_R\,h_R{\boldsymbol{\otimes}}|u|^2\approx f_R |u|^2 -{T_R}\,{t_0^{-1}}\,{\partial_\tau}|u|^2 ]$ with $T_R\approx 3$\,fs \cite{Blow89,GPAbook} [shown in Fig.\,\ref{Raman}(h)], revealing that at lower $\Gamma$, the $P_{\rm cr}$ values deviate significantly from those of the integral model Eq.\,(\ref{eq:1}) (which is the appropriate model in the fs domain). However, both of these models appear to be equivalent once $\Gamma >0.3$. Next, we numerically verify that the sole role of SS parameter has a minimal effect on the switching dynamics, as shown in Fig.\,\ref{Raman}(h). However, for completeness and rigor, we include this perturbation in the coupled NLSE while investigating the combined effects of all perturbations.

\section{IMPACT OF HIGHER-ORDER DISPERSIONS}

\begin{figure}[t]
\centering
\begin{center}   
\includegraphics[width=0.49\textwidth]{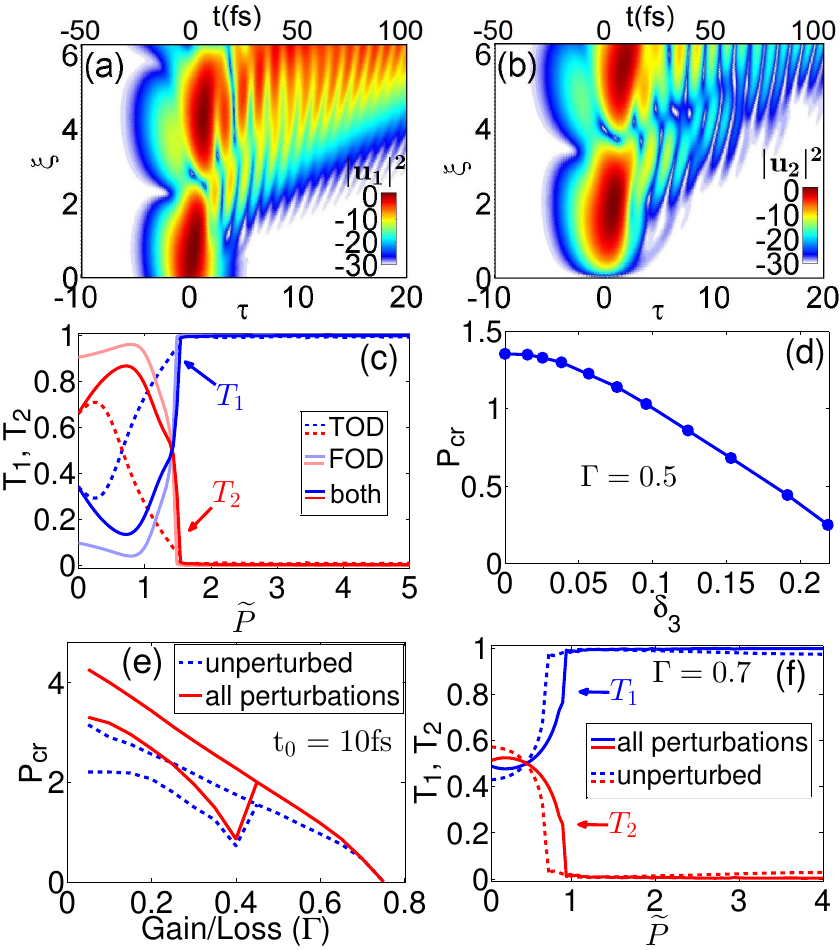}
\caption{(a), (b) Soliton evolution in the two channels of $2\pi$ $\mathcal{PT}$-symmetric coupler under the effect of TOD ($\delta_3\approx 0.153$ of the gain port) for $t_0=5$\,fs with $\widetilde{P}=1$, $\kappa=1$, and $\Gamma=0.5$. (c) Switching dynamics in the presence of TOD only, FOD only ($\delta_4\approx-0.03$), and combined effects for $t_0=5$\,fs. (d) $P_{\rm cr}$ as a function of TOD parameter $\delta_3$ of the gain port with $\kappa=1$. (e) $P_{\rm cr}$ as a function of $\Gamma$ under the combined influence of IRS, SS, and HOD effects, and (f) corresponding switching dynamics} \label{beta3}
\end{center}
\end{figure}

In soliton propagation, HODs are of particular importance and for a given waveguide geometry, the dispersion profile may vary rapidly with the frequency, making HODs more pronounced. While the interplay between Kerr nonlinearity and GVD produces the stable solitonic structure in time and frequency domains, HODs lead to a significant temporal and spectral distortion. More specifically, the soliton sheds energy in the form of dispersive waves (DWs) that produce several discrete spectral peaks \cite{GPAbook,Akhmediev95}. In this process, the temporal distribution of the soliton is affected by the generation of asymmetric (for TOD) and symmetric (for FOD) side-lobes. Here, we consider a $2\pi$ $\mathcal{PT}$-symmetric single-mode fiber coupler, the cross-sectional geometry for different channels and corresponding GVD profiles of which are shown in Figs.\,\ref{Schem}(a) and \ref{Schem}(b). For a realistic scenario, we launch the optical pulse at the gain port considering the individual GVD profiles (second-order GVD, TOD, FOD terms) of the gain and loss fibers, which, at the launching wavelength $\lambda_0=1.55\,\mu$m are calculated as $\beta_{2}\approx-39.051$\,ps$^2$/km, $\beta_{3}\approx0.17959$\,ps$^3$/km, $\beta_{4}\approx-4.8046\times 10^{-4}$\,ps$^4$/km, and $\beta_{2}\approx-18.643$\,ps$^2$/km, $\beta_{3}\approx0.16604$\,ps$^3$/km, $\beta_{4}\approx-4.393\times 10^{-4}$\,ps$^4$/km, respectively. The back and forth temporal evolution of a fundamental soliton between two channels are illustrated in Figs.\,\ref{beta3}(a) and \ref{beta3}(b) in the presence of TOD by solving Eq.\,(\ref{eq:1}) for a $5$\,fs input $sech$ pulse. In both the figures, the presence of side wings at $\tau>0$ indicates the presence of DWs.
%
Here, the zero-dispersion wavelengths are located at $\lambda\approx 1.3\,\mu$m for the gain fiber and $\lambda\approx 1.45\,\mu$m for the loss fiber [can be obtained from the GVD profiles of Fig.\,\ref{Schem}(b)], which are near to the launching wavelength $\lambda_0=1.55\,\mu$m. As a result, relatively stronger dimensionless TOD parameters (which determine the strength and locations of radiation energy), $\delta_3\approx 0.153$ of the gain fiber and $\delta_3\approx 0.297$ of the loss fiber, lead to stronger DW-generation, which significantly impact the switching dynamics. 
Like the impact of IRS, the critical switching power is modified in the presence of HOD terms, as shown in Fig. \ref{beta3}(c). In this scenario, the HOD, in particular the TOD, lowers the $P_{\rm cr}$ to a level much below that of the unperturbed counterpart; however, a decrease in the sharpness of switching is observed. It is worth noting that the consideration of the separate GVD profile in gain and loss channels is critical (which is indeed the practical case), because it results in a lower $P_{\rm cr}$, as opposed to the case with the equal GVD profiles in the two channels that increase the $P_{\rm cr}$. 
For further investigation of the dependence of $P_{\rm cr}$ on the strength of $\delta_3$, we plot $P_{\rm cr}$ as a function of $\delta_3$ in Fig.\,\ref{beta3}(d) for $\Gamma=0.5$ and $\kappa=1$. These figures suggest that by decreasing $t_0$ (increasing the strength of $\delta_3$), we can achieve a very low critical power of switching. However, this perturbation destabilizes the soliton propagation, and to overcome that, IRS plays a crucial role which we discuss in the following.

\section{COMBINED EFFECT OF HGHER-ORDER PERTURBATIONS}
\begin{figure}[t]
\centering
\begin{center}   
\includegraphics[width=0.49\textwidth]{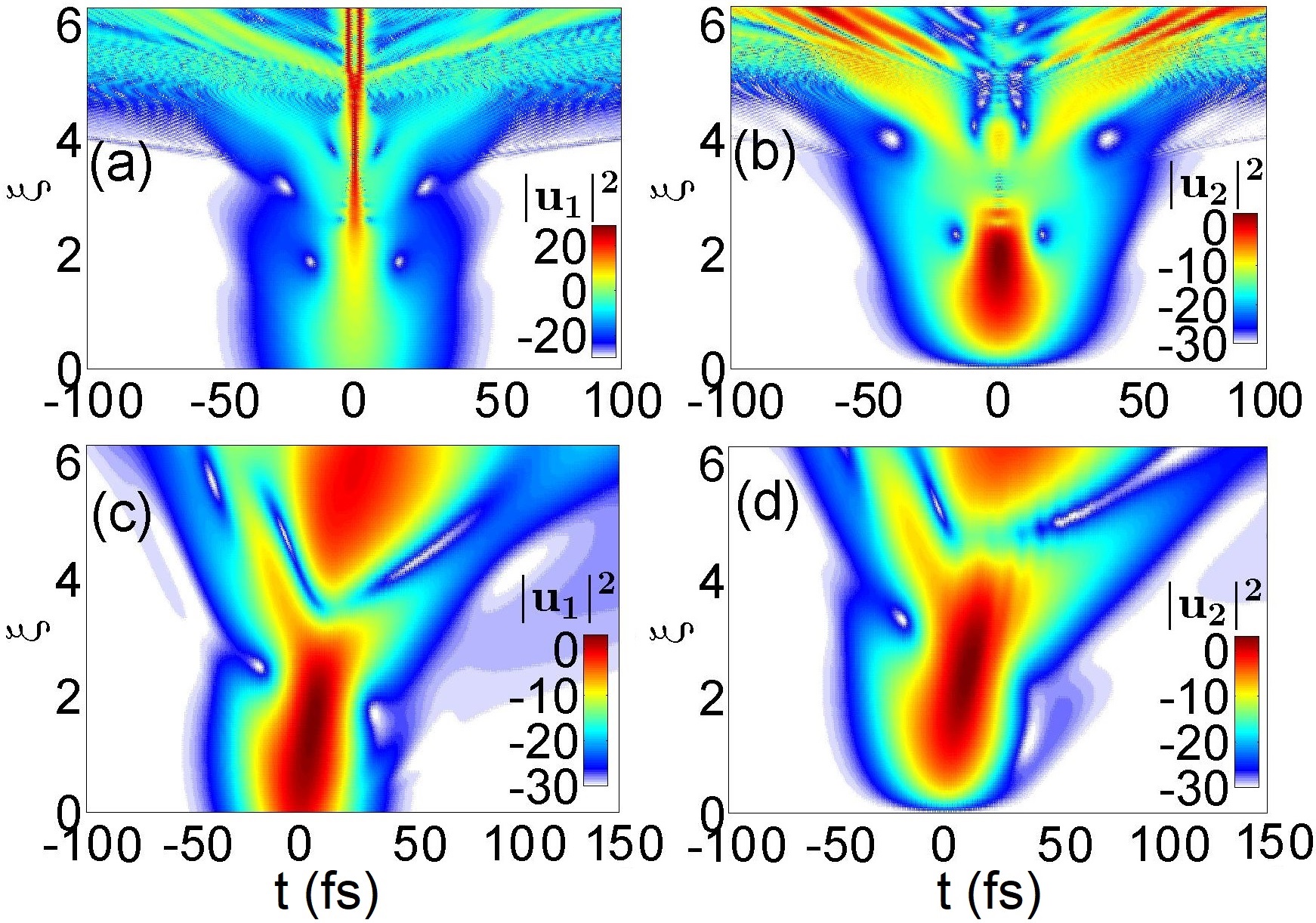}
\caption{Soliton evolution in the two channels of $2\pi$ $\mathcal{PT}$-symmetric coupler in the nonlinear switching domain without perturbations (a), (b), and with all perturbations (c), (d) for $\kappa=\widetilde{P}=1$, $\Gamma=0.65$, and $t_0=10$\,fs.} \label{combined}
\end{center}
\end{figure}
So far, we have discussed the individual perturbative effects of both higher-order nonlinearities (IRS and SS) and HODs on the switching dynamics of a fs pulse in detail. Where the modified parameters of evolving solitons in a single NLSE case \cite{GPAbook} caused by individual perturbation (which we summarize in previous sections) modify the switching dynamics collectively as modes from two channels interact with each other. Now, we analyze the combined effects of the above-stated higher-order perturbations on the switching dynamics. For this, we plot the variation of critical switching power as a function of gain/loss parameter as shown in Fig.\,\ref{beta3}(e). This figure depicts a complete picture of the switching dynamics taking into account all perturbations, demonstrating the parameter range for which the lower critical power exists. Here, although the combined effects of all the perturbations (which is the practical case for the fs soliton in the couplers) increase the critical switching power for lower gain/loss values than that of the unperturbed case (which is limited only in the ps domain), the difference between both the critical switching powers is minimal for higher gain/loss values ($\Gamma>0.6$). After identifying the sweet-spot region for the low power switching in the fs domain, we next figure out how the efficient switching takes place in the high $\Gamma$ region. For that, we plot the transmission energy for a range of input pump power at $\Gamma=0.7$ in Fig. \ref{beta3}(f). Although the critical power remains the same as in the unperturbed case ($P_{\rm cr}\approx 0.42$), the switching steepness improves relative to the perturbed cases. Also, the nonlinear switching ($\widetilde{P}>P_{\rm cr}$) can be achieved with fundamental soliton ($\widetilde{P}=1$) or lower ($P_{\rm cr}<\widetilde{P}<1$) as input,  resulting in more robust spatiotemporal evolution under perturbations. More importantly, the stability of solitons and switching efficiency improve substantially in the fs regime when compared to the unperturbed case (which is practically limited in ps time scale). Therefore, we can achieve a very low critical power of switching for fs solitons with a complete energy transfer at a relatively higher gain/loss value.  
\begin{figure}[t]
\centering
\begin{center}   
\includegraphics[width=0.49\textwidth]{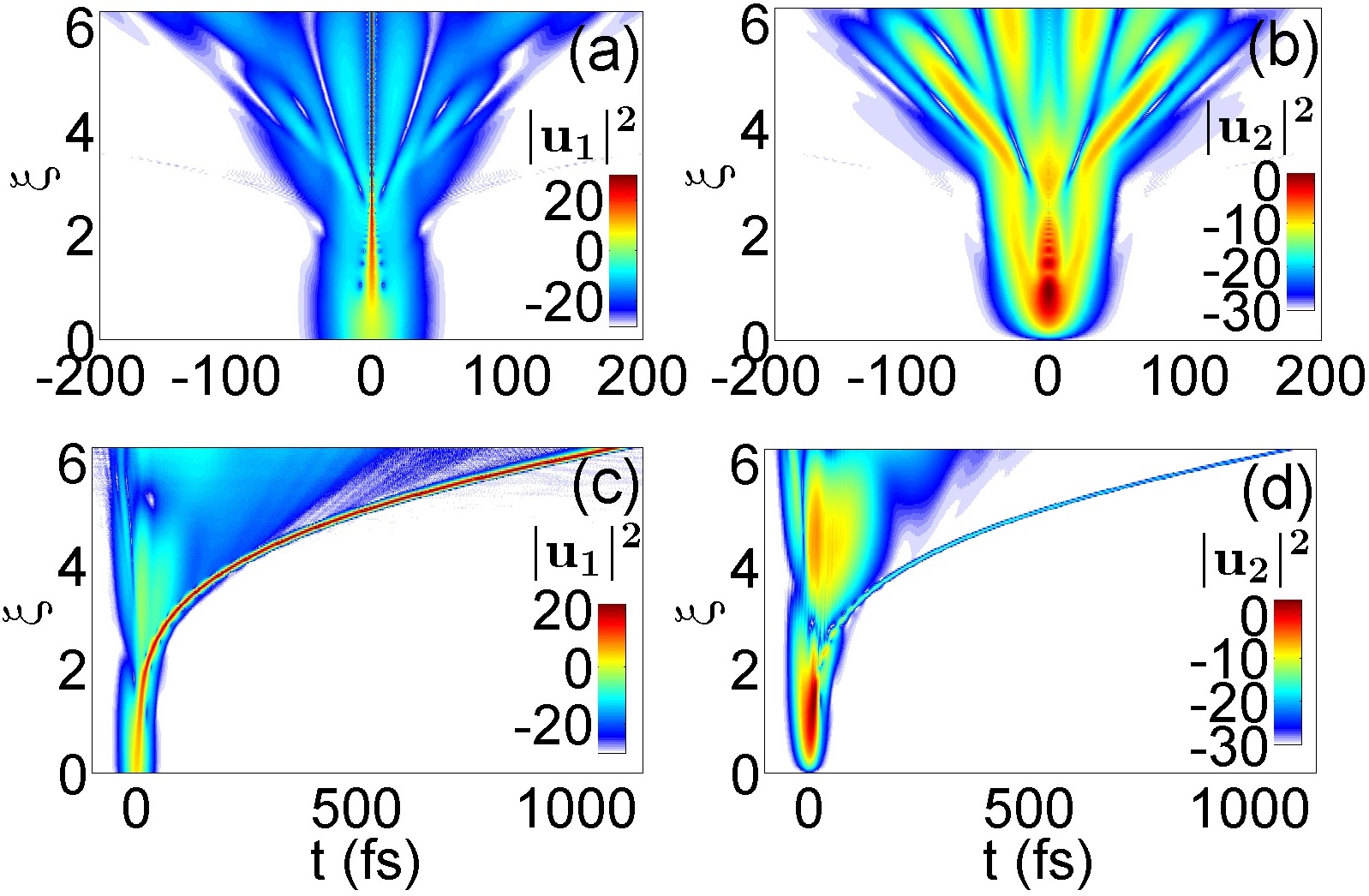}
\caption{Higher-order soliton evolution in the two channels of $2\pi$ $\mathcal{PT}$-symmetric coupler in the nonlinear switching domain without perturbations (a), (b), and with all perturbations (c), (d) for $\kappa=1$, $\Gamma=0.5$, $\widetilde{P}=2.25$, and $t_0=10$\,fs.} \label{nonlinear}
\end{center}
\end{figure}

To validate the fact, we plot the temporal evolution of a fundamental soliton between the two channels, as shown in Figs.\,\ref{combined}(a) and \ref{combined}(b) (without perturbations) and Figs.\,\ref{combined}(c) and \ref{combined}(d) (combination of all the perturbations). These plots demonstrate that with a high gain/loss value, the combined effects of perturbations stabilize the pulse evolution when compared to the chaotic behavior of unperturbed evolutions shown in the top panels of Fig. \ref{combined}. This could be attributed to the fact that IRS and HOD have opposite impacts on the spectral power (IRS tries to redshift the spectrum while HOD, on the other hand, tries to blueshift the spectrum in the form of DWs) manifesting in the self-organizing power flow so as to produce a more stable pulse evolution. We also numerically confirm that, when $P_0=2.25$ (shown in Fig.\,\ref{nonlinear}) which indicates the nonlinear switching domain, the overall effect of perturbations preserves the solitonic nature and assists in focusing the pulse energy from spreading away in contrast to the case of an unstable and chaotic dynamics), which is even not possible when the perturbative effects function alone. It may be useful to note that, while this work convincingly demonstrates the  role of HODs and IRS on the stabilization of unstable solitons, more insights may be obtained by semi-analytical studies such as Lagrange's variational analysis. We have attempted such a rigorous variational study recently \cite{Sahoo-arxiv}. Overall, our analyses show that the practical scenario of the fs domain yields the best results in terms of ultrafast, efficient, and low power switching at relatively high $\Gamma$ values, which is usually considered to be the chaotic one as this condition may lead to spectral singularity \cite{borisl} whenever the value of $\Gamma$ reaches near the value of $\kappa$.

\section{CONCLUSIONS}

To conclude, we have demonstrated theoretically that while in the conventional fiber coupler fs soliton steering is hard to realize as it takes the higher value of critical pump power to steer the pulse due to the various pertubative effects, in a $\mathcal{PT}$-symmetric fiber coupler, these perturbations rather assist in the efficient soliton steering, in the presence of high gain/loss. This work may open up a plethora of applications and studies related to the soliton steering and switching using $\mathcal{PT}$-symmetric fiber couplers in a high loss/gain regime, paving the way for efficient femtosecond all-optical switching devices.


\begin{thebibliography}{1}

\newcommand{\enquote}[1]{``#1''}

\bibitem{Sasikala18}
V. Sasikala and K. Chitra, {\it ``All optical switching and associated technologies: a review,"} J. Opt. {\bf 47}, 307–317 (2018).

\bibitem{friberg1987ultrafast}
S. R. Friberg, Y. Silberberg, M. K. Oliver, et al., {\it ``Ultrafast all-optical switching in a dual-core fiber nonlinear coupler,"} Appl. Phys. Lett. {\bf 51}, 1135 (1987).

\bibitem{friberg1988femotosecond}
S. R. Friberg, A. M. Weiner, Y. Silberberg, B. G. Sfez, and P. S. Smith, {\it ``Femotosecond switching in a dual-core-fiber nonlinear coupler,"} Opt. Lett. {\bf 13}, 904-906 (1988)

\bibitem{Trillo88}
S. Trillo, S. Wabnitz, E. Wright, and G. Stegeman, {\it ``Soliton switching in fiber nonlinear directional couplers,"} Opt. Lett. {\bf 13}, 672-674 (1988).

\bibitem{Ruter10}
C. E. Rüter, K. G. Makris, R. El-Ganainy, D. N. Christodoulides, M. Segev, and D. Kip, {\it ``Observation of parity–time symmetry in optics."} Nat. Phys. {\bf 6}, 192-195 (2010).

\bibitem{El-Ganainy18}
R. El-Ganainy, K. G. Makris, M. Khajavikhan, Z. H. Musslimani, S. Rotter, and D. N. Christodoulides, {\it ``Non-Hermitian physics and PT symmetry,"} Nat. Phys. {\bf 14}, 11-19 (2018).

\bibitem{suchkov2016nonlinear}
S. V. Suchkov, A. A. Sukhorukov, J. Huang, S. V. Dmitriev, C. Lee, and Y. S. Kivshar, {\it ``Nonlinear switching and solitons in PT-symmetric photonic systems,"} Laser and Photonics Rev. {\bf 10}, 177-213 (2016).

\bibitem{suneera2019switching}
T. P. Suneera and P. A. Subha, {\it ``Switching dynamics in parity-time symmetric coupler with nonlocal nonlinearity having transverse potentials,"} J. Modern Opt. {\bf 66}, 1528-1533 (2019).

\bibitem{sukhorukov2010nonlinear}
A. A. Sukhorukov, Z. Xu, and Y. S. Kivshar, {\it ``Nonlinear suppression of time reversals in $\mathcal{PT}$-symmetric optical couplers,"}  Phys. Rev. A {\bf 82}, 043818 (2010).

\bibitem{ramezani2010unidirectional}
H. Ramezani, T. Kottos, R. El-Ganainy, and D. N. Christodoulides, {\it ``Unidirectional nonlinear $\mathcal{PT}$-symmetric optical structures,"} Phys. Rev. A {\bf 82}, 043803 (2010).

\bibitem{dmitriev2011scattering}
S. V. Dmitriev, S. V. Suchkov, A. A. Sukhorukov, and Y. S. Kivshar, {\it ``Scattering of linear and nonlinear waves in a waveguide array with a $\mathrm{PT}$-symmetric defect,"} Phys. Rev. A {\bf 84}, 013833 (2011).

\bibitem{Driben11}
R. Driben and B. A. Malomed, {\it ``Stability of solitons in parity-time-symmetric couplers,"} Opt. Lett. 36, 4323-4325 (2011).

\bibitem{alexeeva2012optical}
N. V. Alexeeva, I. V. Barashenkov, A. A. Sukhorukov, and Y. S. Kivshar, {\it ``Optical solitons in $\mathcal{PT}$-symmetric nonlinear couplers with gain and loss,"} Phys. Rev. A {\bf 85}, 063837 (2012).

\bibitem{Fan19}
Z. Fan and B. A. Malomed, {\it ``Dynamical control of solitons in a parity-time-symmetric coupler by periodic management,"} Commun. Nonlinear. Sci. Numer. Simul. {\bf 79}, 104906 (2019).

\bibitem{Govindarajan19}
A. Govindarajan, A. K. Sarma, and M. Lakshmanan, {\it ``Tailoring $\mathcal{PT}$-symmetric soliton switch,"} Opt. Lett. {\bf 44}, 663-666 (2019).

\bibitem{Blow89}
K. J. Blow and D. Wood, {\it ``Theoretical description of transient stimulated Raman scattering in optical fibers,"} IEEE J. Quantum Electron. {\bf 25}, 2665 (1989).

\bibitem{GPAbook} 
G. P. Agrawal, {\textit{Nonlinear Fiber Optics}}, 5th ed. (Academic, New York, 2013).

\bibitem{Sahoo22} 
A. Sahoo, D. K. Mahato, A. Govindarajan, and A. K. Sarma, {\it ``Bistable soliton switching dynamics in a $\mathcal{PT}$-symmetric coupler with saturable nonlinearity,"} Phys. Rev. A {\bf 105}, 063503 (2022).

\bibitem{Malomed97} 
B. A. Malomed, I. M. Skinner, and R. S. Tasgal, {\it ``Solitons in a nonlinear optical coupler in the presence of the Raman effect,"} Opt. Comm. {\bf 139}, 247-251 (1997).

\bibitem{Myslinski97}
P. Myslinski, D. Nguyen, and J. Chrostowski, {\it ``Effects of concentration on the performance of erbium-doped fiber amplifiers,"} J. Lightwave Technol. {\bf 15}, 112 (1997).

\bibitem{Sahoo21}
A. Sahoo and A. K. Sarma, {\it ``Microresonator dynamics with frequency-dependent Kerr nonlinearity,"} Phys. Rev. A {\bf 104}, 023513 (2021).

\bibitem{Lin06} 
Q. Lin and G. P. Agrawal, {\it ``Raman response function for silica fibers,"} Opt. Lett. {\bf 31}, 3086-3088 (2006).

\bibitem{Gordon86} 
J. P. Gordon, {\it ``Theory of the soliton self-frequency shift,"} Opt. Lett. {\bf 11}, 662-664 (1986).

\bibitem{Akhmediev95} 
N. Akhmediev and M. Karlsson, {\it ``Cherenkov radiation emitted by solitons in optical fibers,"} Phys. Rev. A {\bf 51}, 2602 (1995).

\bibitem{Sahoo-arxiv} 
A. Sahoo and A. K. Sarma,{\it``Variational approach to study solitary waves in $\mathcal{PT}$-symmetric nonlinear couplers," } arXiv:2206.09468.

\bibitem{borisl}
R. Driben and B. A. Malomed, {\it ``Stabilization of solitons in PT models with supersymmetry by periodic management,"} Europhys. Lett. {\bf 96}, 51001 (2011).



	
\end{thebibliography}
\end{document}